# Methods for an Electron Emission Digital Twin

Salvador Barranco Cárceles[1,4*], Veronika Zadin[2], Steve Wells[3]
Aquila Mavalankar[3], Ian Underwood[4], Anthony Ayari[1]

[1]Institute Lumière Matière, Université Claude Bernard Lyon 1, Villeurbanne, France
[2]Instute of Technology., Tartu, Estonia
[3]Adaptix Imaging Ltd., Oxford, England
[4]School of Engineering, University of Edinburgh, Edinburgh, Scotland
[*]Corresponding author: salvador.barranco-carceles@univ-lyon1.fr, s.barranco.carceles@gmail.com

## ABSTRACT

The effective design and operation of electron emitters is the core of critical technologies such as high-resolution electron imaging and spectroscopy or X-ray production for medical imaging. Despite 100 years of theoretical development in thermo- and field-electron emission models, the analysis of experimental data and design of electron emitters remains an art more than a science. This is due to the many processes that are involved in electron emission, which result in an extremely complex phenomenon. Here we describe and develop the Methods for an Electron Emission Digital Twin (MEEDiT), which integrates state-of-the-art thermo-field electron emission models and experimental data characterisation. By applying MEEDiT to silicon electron emitters, we demonstrate an approach that bridges the gap between simple experimental measurements and 'hidden' physical quantities like temperature and field enhancement. MEEDiT provides the physical consistency of a 3D simulation with the speed of a neural network, enabling resource-effective, real-time characterization and the extraction of critical data that is otherwise inaccessible during operation.

*Key works:* Electron Emission, PINN, Digital Twin

## I. Introduction

The emission of electrons from solids has been at the centre of several scientific and industrial revolutions. The Photoelectric Effect [1], [2] triggered the era of the quantum framework for physics and enabled the development of a whole family of photoemission spectroscopy techniques [3], [4], [5], [6]. Thermionic Electron Emission [7] allowed early studies of work functions of solids and surface potentials [8], [9] as well as underpinned the development of electronic devices such as the vacuum amplifier [10]. Field Electron Emission [11] provided experimental proof of quantum tunnelling and established the path for technologies such as the Field Ion Microscope (first images of atoms) [12], [13] and the Transmission Electron Microscope [14]. The common aspect of these phenomena and technologies is their sensitivity to the changes in the surface conditions; giving, thus, birth to the field of surface physics.

The fact of having a hard stop to the periodicity of the solid at the interface with vacuum leads to variation of the interatomic distance in the boundary layer. Consequently, it results in the appearance of the work function [15], variations of the density of states [16], and the appearance of surface states [17]. In addition, dimensionality, which may range from a flat surface several millimetres long to a curved surface a few nanometres short, causes additional modifications of the surface potential [18], the available energy states [19], or the mathematical treatment of the quantum problem [20].

The existence of a surface leaves place for parasitic interactions between the *pristine* surface of the solid and the trace gases present in vacuum systems [21]. These parasitic interactions degrade the level of electron emission over time, enforce high vacuum levels for device operation [22], and ultimately results in device failure in the form of vacuum break down.

These surface induced phenomena and parasitic effects make, even after 100 years of research, the study [23], commercialisation [24], and operation [25], [26] of electron emission devices rather challenging given their complex and interdependent nature.

The recent surge of Machine Learning (ML) and Physics Informed Neural Networks (PINN) [27], [28], [29], [30] have been applied to the prediction of vacuum arcs in rf cavities [31], in which an increase of the vacuum pressure was correlated to the vacuum arc event. ML has also been used to predict the emitted current from flat cathodes [32]. While pioneering, these approaches still fail to provide fine details about the complex phenomena of electron emission from surfaces under high electric fields (e.g., temperature, field enhancement, voltage drop, etc.).

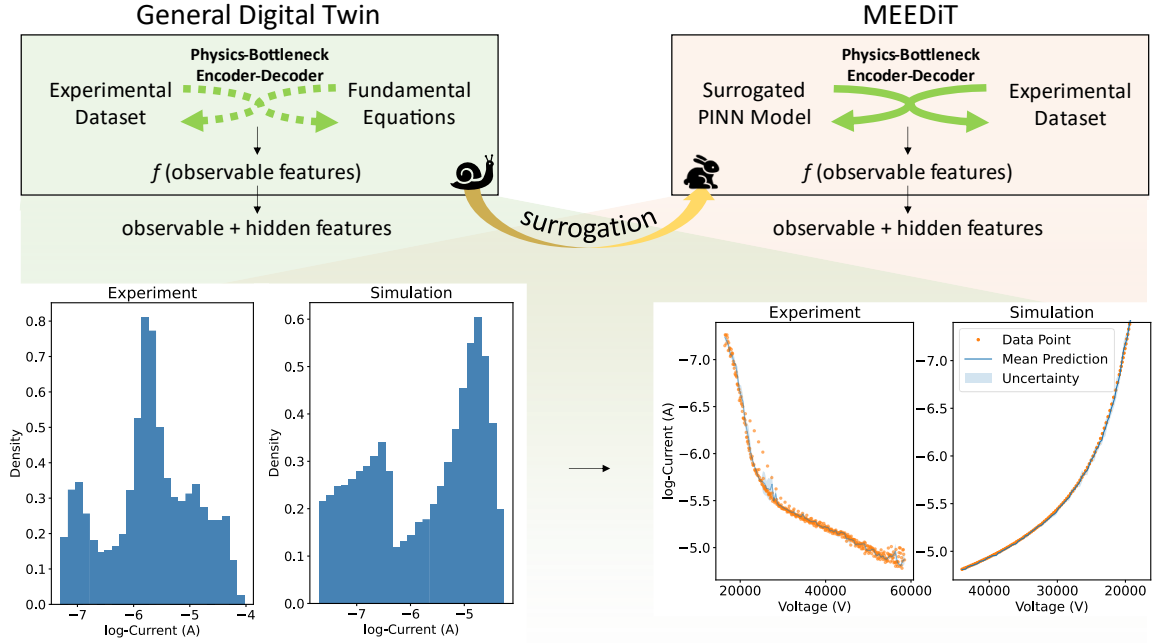

Figure 1: Conceptualisation of MEEDiT and the surrogation process. The top left shows the general concept of a digital twin, based on physics-bottleneck encoder-decoder architecture, where the fundamental equations (e.g., Fowler-Nordheim field emission equations) are used to build the digital twin and to extract *hidden features* (e.g., field enhancement factor). Given that the 1D equations can not capture the complex behaviour of 3D objects, a PINN is used to develop a surrogated model that replaces those fundamental equations (top right). To create the surrogated model, several 3D simulations are run over a parameter space rich enough to capture the behaviour of the real object. The bottom left histograms show that the model by Barranco-Cárceles *et al.* [37] replicates the main current characteristic of real field emitters. The bottom-right graphs show that a PINN can be trained in both the experimental and synthetic dataset to be later used as a surrogate for MEEDiT; allowing the implementation of the surrogate for accurate and fast computation of the emission characteristics.

In this work we introduce the concept of a Digital Twin (MEEDiT) as a framework to bridge the gap between theory, experiments and applications for electron emission. We envision MEEDiT as a versatile tool for both scholarly research and for industrial device design with on-the-fly operation and prediction capacity.

## II. Results

### The Surrogated Theoretical Model

The primary objective of MEEDiT is to estimate the working conditions of an electron emitter, with limited information about the system (e.g., local pressure unknown, local heat unknown; but general geometry and I-V known), so the emitter can be operated and conditioned effectively. A solution for such a problem can be found in the spirit of Kalman Filters [33], where a set of noisy/approximated sensor measurements are processed through a set of linear equations to filter out noise and estimate the *true* state of the system. Certainly, for an electron emitter this can be done by applying any of the well-known analytical solutions [34], [35], [36]. However, analytical solutions are limited to 1D approximations and cannot capture the complex temperature or potential gradients paramount to describing a realistic electron emitter. While full 3D Finite Element Models [37] or ab-initio approaches [38], [39], [40] offer higher fidelity, their computational cost is prohibitive for real-time inference or large-scale parameter sweeps (Fig. 1 top left).

Our surrogated approach overcomes this bottleneck by training a Probabilistic Physics-Informed Neural Network (PINN) that acts as a surrogated model [41]. This surrogate is pre-trained on high-fidelity synthetic data to replicate the underlying physical laws governing electron emission (Fig. 1 bottom left and right). By emulating the complex analytical equations with a neural network, the model serves as a high-speed computational core for the development of MEEDiT. This allows us to infer hidden physical quantities (such as temperature and field enhancement) from experimental measurements with the speed of a neural network and the physical consistency of a 3D simulation (Fig. 1 top right).

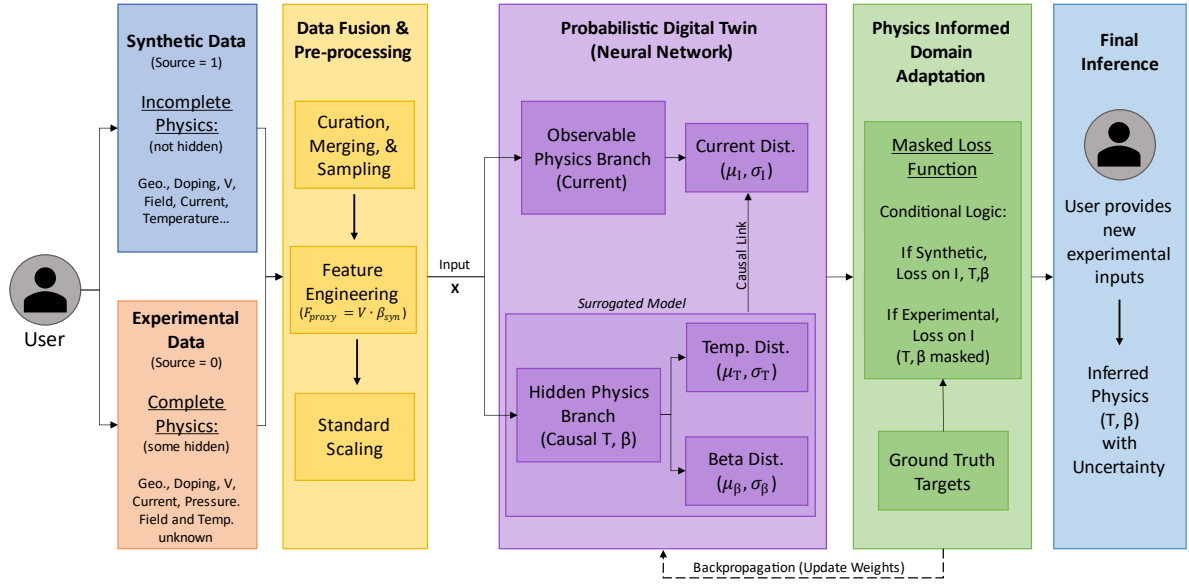

Figure 2: Schematic of the data workflow of MEEDiT. The user provides the synthetic (dark blue block) and the experimental (orange block) data sets that contain information about the geometry, electric properties, voltages, currents… Those two datasets are cleaned to removed abrupt changes (e.g., an emitter is destroyed) and merged into one with their corresponding labels (source 1 or 0). New features are added to enrich the learning process: an estimation of the field ($F_{proxy} = \beta V$) can be added from knowing the potential and geometry since $I \propto (\beta V)^2 e^{-\phi^{3/2}/\beta V}$. Non-linear data (e.g., current or pressure change) is rescaled in into logarithmic scaled and the data is normalised to prevent resolution error propagation (yellow block). Based on the passed data and the causal link that exists between the emitted current and the temperature and field at the emission site, an estimation and uncertainty is provided (purple block). The features mean and std. distribution are passed to a loss function that discriminates between synthetic and experimental data and compare them with a "ground truth target" (testing data) (green block). The new weights are backpropagated until the convergence criteria is met. The weights and scaling factors are saved for later use in analysis or control (light blue block).

To develop the surrogated model, we used the model published by Barranco-Cárceles *et al.* [37] and a parameter domain (tip radius, emitter angle and length, doping, …) that covers a wide range of the experimental data. The data from the simulations, thereafter synthetic data, capture the main behaviour of the experimental data (as can be seen in the current histograms, Figure 1 bottom left, where two peaks are visible). The synthetic data also contains information about the temperature, electric field, and field enhancement factor; which are inaccessible from experiments.

Once the synthetic data is produced, we executed a proof of concept and data quality control by training a Random Forest ML (as it gave satisfactory results in previous studies [31], [32]) to verify its suitability for training MEEDiT. The Random Forest, using 65% of data available for training, gave satisfactory results in learning and predicting the behaviour of both experimental and simulated emitters (Fig. 1 bottom left). By subtracting a linear fit from the comparison between predicted and validation data, we obtained a Gaussian distribution of residuals (not shown). This confirms the successful implementation of the surrogate model to replace the physics model based in equations.

*From Theory Models to Digital Twins*

The main goal of MEEDiT v.01 involves the estimation of the emitted current, the temperature of the emitter, and the field enhancement factor (β), given a set of experimental parameters such as the geometry of the emitter, the material of the emitter, the applied potential, and the pressure change. Based on the surrogated approach detailed above, we construct the architecture of the Digital Twin as follows.

The user provides two data sets (Fig 2 left) which are labelled: "Source 1" and "Source 0". The synthetic data (source = 1) is generated from accurate 3D simulations and acts as the "physics instructor". The experimental data (source = 0) is incomplete (e.g., temperature not included) but acts as "ground truth" to anchor the training to reality. These two data sets are merged into a single hybrid dataset (Fig 2 yellow block). To enrich the learning phase, engineered features were added to provide the model with explicit physical relationships. This streamlines the learning process by bypassing the need for the network to infer these dependences independently. The data is then standardized (Z-score Scaling) to ensure numerical stability during gradient descent and send as an array to the neural network (see "Input X" in Fig. 2).

The central block of the diagram (Fig 2 purple block) represents the neural architecture of MEEDiT. We employ a Physics-Bottleneck Encoder-Decoder design that mirrors the causal chain of electron emission. The vector X feeds into the Hidden Physics Branch (the Encoder), a sub-network that acts as the Surrogate. Its objective is to map the geometry and operating conditions of the emitter to its latent physical state, so it learns how the 3D structure of the emitter affects the field enhancement and the temperature. These predictions are then passed into the Current Branch (the Decoder). The Causal Link creates the Physics Bottleneck.

The model cannot predict an emitted current value without estimating plausible values for the enhancement factor and the temperature. This enforces geometry constrains into temperature and enhancement factor, which in turn determines the emission current for a given voltage. The Causal Link can be thought as a loop to obtain self-consistency. The outputs from this block are not scalar values, but the Gaussian distributions for every targeted variable (as in a Kalman Filter). This transforms MEEDiT into a probabilistic inference engine that provides confidence interval for every prediction.

The Green Block (Fig. 2) describes the optimisation engine, which is driven by a custom Masked Negative Log-Likelihood loss function. This function dynamically alters its learning objective based on the data source, enabling it to solve the Inverse Problem. When the model encounters synthetic data (source = 1) the loss mask is fully activated; thus, minimising the error for the estimators. Basically, this forces the Hidden Physics Branch to accurately replicate the theorical physics for the surrogate so that the estimations are physically sound. When the model encounters experimental data (source = 0), the error is only calculated for the emission current. The model suggests values for the field enhancement factor and for the temperature to estimate the experimental current. The error between the estimated and experimental current generates a gradient. This gradient is then propagated backwards to update the weights of the Surrogated Encoder. This is repeated until the model meets the convergence criteria and a solution is found.

The Final Interface (Fig. 2 grey block) presents the trained Digital Twin, with calibrated internal physics representing the experimental reality. A user can input new geometries and experimental features (e.g., voltage, pressure, doping…) and the model, acting as a probabilistic inverse solver, returns estimations and uncertainties of the temperature and the field enhancement factor.

*Show Case Example*

A classic example of an electron emitter is a tip that, under high electric field, emits electrons and due to Joule and Nottingham heating undergoes thermal runaway [42]. Additionally, because of the presence of a potential gradient inside the emitter and thus along its surface, there is an accumulation of foreign species at the top of the tip that heavily modifies the electron emission [22], [43]. These present both a fundamental challenge in the study of surfaces [44] and a technological limit in research [45] and application [46]. Having insight into the downstream events that lead to thermal runaway and/or surface modification enables the development and application of strategies for mitigation.

While MEEDiT is compatible any electron emission method (e.g., photo- or field- emission), we have implemented it here to model thermo-field electron emission from semiconductors as primary demonstration of our Digital Twin. We have chosen semiconductors because they exhibit a more complex emission nature than metallic emitters (e.g., band bending, surface states, doping distribution, internal potential drop, non-linear emission, etc.) and thus we expect to demonstrate the power of the methods described here.

The training data consists of two sets: an experimental dataset and a synthetic dataset. The experimental dataset holds information of some 500 different emitters (tip radius, angle, height, and doping) that have undergone several I-V runs. The geometric parameters are extracted by Scanning Electron Microscopy, the doping is known from the manufacturer's data sheet, and the I-V characteristics are obtained by averaging several cycles of I-V with a sequence of progressively higher potentials until the emitters fails. The data recorded beyond emitter failure, as recognised by a sudden non-smooth change in the I-V, is discarded for this version of MEEDiT since the geometry becomes unknown (we are already developing MEEDiT-v02 that addresses this aspect). Similarly, as high voltage data is discarded, the low voltage data, where noise from the current amplifier is recorded, is also removed to prevent spurious results. The features that are passed from the experimental dataset to MEEDiT are: doping, tip radius, angle, height, voltage, current, pressure change. The passed features can be, however, tailored to fit the needs of other experimental setups (e.g., CNT chirality).

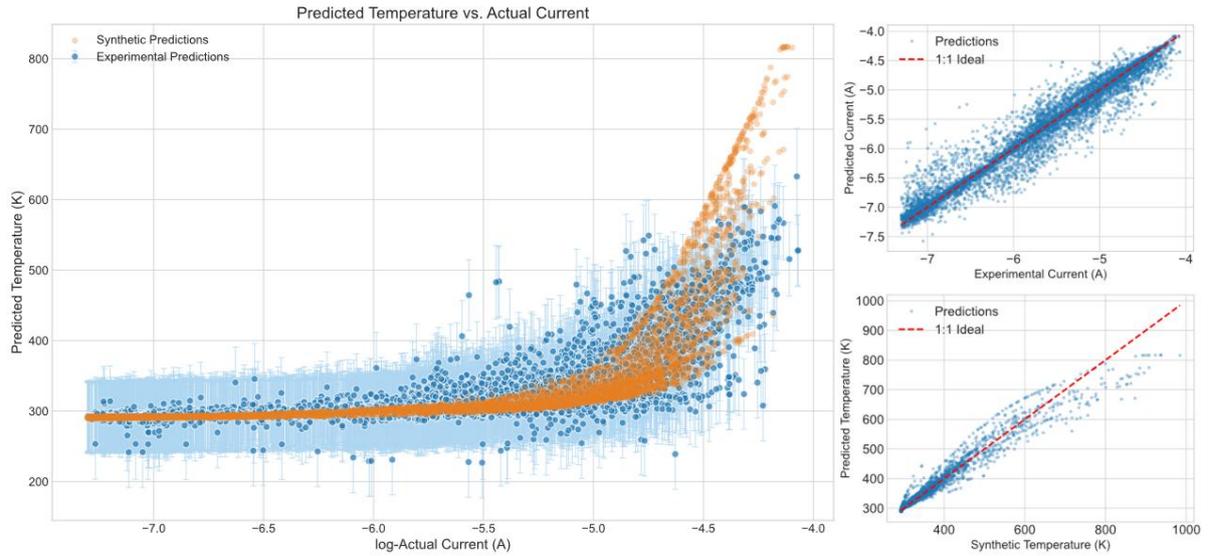

Figure 3: Predictions made by MEEDiT versus experimental and synthetic data. The left-hand side graph shows the temperature as a function of the emitted electron current from the simulated data (orange) and predicted data (blue). Both curves follow the same trend, but it can be seen that the simulations overestimate maximum temperature that an emitter can reach before failure. This observation is made obvious in the deviation from the 1:1 in the bottom right graph. The top right graph shows a strong correlation between the experimental and the predicted current, as expected from the accurate 3D model used to produce the synthetic data.

The synthetic data is obtained from high-fidelity 3D models [37] where the geometry and electrical properties of the emitters are replicated. It is recommended to run a sensitivity analysis to identify the most important inputs of the simulation and, thus, enhance the simulations speed. In our particular case, the most sensitive parameter is the tip radius and thus we fine-tuned the simulations for a denser grid corresponding to the radius. The resulting current distribution shows two peaks at currents close to the experimental data, showing the good agreement of the 3D model and yet with subtle differences (Fig. 1 bottom left). The simulation features that passed onto MEEDiT are: doping, radius, angle, height, $\beta$, voltage, current, temperature.

The features that are passed to MEEDiT bundle together the main parameters that characterise the system. However, the learning process can be accelerated and made more effective by adding *engineered* features that represents the system under study. For instance, we have included a "field_proxy" feature that relates the applied voltage to $\beta$, from which the model can learn the exponential dependence between electric field and emitted current, as well as the non-linear dependence between $\beta$ and voltage that semiconducting emitters present. Some other user might be inclined to add extra proxy features to enhance the learning (e.g., inverse of the field (voltage) and the log of the current over the field (voltage) squared or the relationship between laser power and emitted electrons). In our case one proxy feature was sufficient. In addition to adding extra features, in order to enhance the model robustness and ensure numerical stability, we applied a logarithmic transformation to skewed feature distributions (e.g., emitted current), effectively linearizing their relationship with the target variable. Subsequently, all features were z-score normalized to achieve a zero-mean and unit-variance distribution, preventing objective function bias toward high-magnitude variables.

Figure 3 shows the outputs of MEEDiT after training. The graph on the left-hand side shows temperature as a function of the emitted current T(I). In orange the synthetic data from the model and in blue the estimated T(I) from MEEDiT. The estimated data follows the expected trend, increasing temperature for increasing current, but it falls short in determining the upper limit of high temperature if compared with the synthetic data. This is not a bug of the model but evidence that MEEDiT is estimating the temperature realistically. In static simulations, the model allows emission to be pushed to a point that temperature is beyond the melting point of the material. In an experiment the tip of the emitter is subjected to electrostatic forces that are constantly pulling the tip. As the temperature increases, the atoms at the tip become loose from their atomic positions which, in the presence of the high pulling forces, results in a inelastic deformation of the tip that finalises in vacuum breakdown [42]. Thus, the emitter might not need to reach the melting point to undergo thermal runaway, but a threshold temperature that is approximately 30% of their bulk melting point [44], which is ~500 K as estimated by MEEDiT. The deviation from the ideal static simulation to the Digital Twin estimation can be clearly seen in Fig 3 (bottom right) where the synthetic and estimated data do not fall on the 1-1 line for the temperature. It does, however, for the comparison between predicted and experimentally measured current (Fig. 3 top right) proving the capacity of the Digital Twin to estimate the current.

## III. Discussion

A general theoretical model that includes all the physical phenomena involved in electron emission from surfaces remains a challenging and elusive task due to the broad range of scales (e.g., angstroms to millimetres, picoamps to milliamps, etc.) and the non-linear relationship among them. The Digital Twin approach presents a new perspective that shifts from highly detailed models to approximate models that, in conjunction with experimental data, better capture the behaviour of the system as a whole. By applying MEEDiT to silicon electron emitters, we demonstrate that this approach can bridge the gap between simple experimental measurements and 'hidden' physical quantities like temperature and field enhancement. MEEDiT provides the physical consistency of a 3D simulation with the speed of a neural network, enabling cost-effective, real-time characterization and the extraction of critical data that is otherwise inaccessible during operation

However, MEEDiT v.01 presents several limitations. First, the current implementation does not take into account dynamically changing surfaces under steady state conditions (e.g., a field emitter that is left at constant voltage, but the emitted current changes because of modifications of the surface conditions). A potential solution is to add a time dimension to the model, that is governed by an extra surrogate based on molecular dynamics models to capture the surface diffusing under high fields. Second, our Digital Twin does not take into account the electron trajectories. Currently, once an electron is emitted it is considered *gone* (with the exception of its contribution to the pressure change due to electron induce degassing of the anode). Taking into account electron trajectories would be of interest for the electron microscopy [47] and plasma initiation communities [48]. The third limitation, is that MEEDiT v.01 does not go beyond the point of emitter failure. This is mainly due to vast zoology of topographies that results after a vacuum break down event. Further experimental and theoretical work are required to shine light on these particular topics and will be addressed in future versions of MEEDiT.

Despite its current limitations, MEEDiT has proven to be a versatile tool for the characterization and operation of electron emission systems. We envision this probabilistic approach as a powerful framework for studying surfaces under high electric fields—a domain that extends beyond classical field emission into modern investigations of molecules on surfaces [49] and light-matter interactions [50]. By incorporating more granular data structures, such as total energy distributions, and refining our noise models, we expect MEEDiT to evolve into a fully realized Digital Twin. As such, it will provide real-time data interpretation and predictive analytics at the cutting edge of experimental physics.

*ACKOWLEDGEMENTS*
This work has been supported by the Alan Turing Institute grant ATI/TES/2023-09, by the Estonian Research Council grants PRG2675, TARISTU24-TK10 and TEM-TA23, and by the Agence National de la Recherche grant ANR-22-CE09-0021.

*AUTHOR DECLARATIONS*
**No conflict of interest.** The authors have no conflicts of interest to disclose.